\documentclass[aps,twocolumn,a4paper,pra,superscriptaddress,floatfix]{revtex4}%
\usepackage{amssymb}
\usepackage{amsmath}
\usepackage{amsfonts}
\usepackage{graphicx}
\usepackage{graphics}

\def\P{\mathrm{P}}    %plasma
\def\C{\mathrm{C}}    %corrugation
\def\PP{\mathrm{PP}}  %plane-plane
\def\PS{\mathrm{PS}}  %plane-sphere
\def\TE{\mathrm{TE}}  
\def\TM{\mathrm{TM}}  
\def\lat{\mathrm{lat}}  
\def\br{\mathbf{r}}
\def\bk{\mathbf{k}}
\def\dd{\mathrm{d}}
\def\kP{k_\mathrm{P}}    %plasma
\def\kC{k_\mathrm{C}}    %corrugation
\def\corr{\mathrm{corr}}    
\def\PFA{\mathrm{PFA}}

\begin{document}
\title{The lateral Casimir force beyond the proximity force approximation~:
a nontrivial interplay between geometry and quantum vacuum}
%\runningtitle{The lateral Casimir force beyond the proximity force approximation}
\author{Robson B. Rodrigues}
\author{Paulo A. Maia Neto}
\affiliation{Instituto de F\'{\i}sica, UFRJ, 
CP 68528,   Rio de Janeiro,  RJ, 21941-972, Brazil}
\author{Astrid Lambrecht}
\author{Serge Reynaud}
\affiliation{Laboratoire Kastler Brossel,
CNRS, ENS, Universit\'e Pierre et Marie Curie case 74,
Campus Jussieu, F-75252 Paris Cedex 05, France}

\date{\today}

\begin{abstract}
The lateral Casimir force between two corrugated metallic plates makes possible 
a study of the nontrivial interplay of geometry and Casimir effect appearing
beyond the regime of validity of the Proximity Force Approximation (PFA).
Quantitative evaluations can be obtained by using scattering theory in 
a perturbative expansion valid when the corrugation amplitudes are smaller
than the three other length scales: the mean separation distance $L$ of the plates,
the corrugation period $\lambda_\C$ and the plasma wavelength $\lambda_\P$. 
Within this perturbative expansion, evaluations are obtained for arbitrary
relative values of $L$, $\lambda_\C$ and $\lambda_\P$ while limiting cases,
some of them already known, are recovered when these values obey some
specific orderings. 
The consequence of these results for comparison with existing experiments
is discussed in the end of the paper. 
\end{abstract}
\maketitle

\section{Introduction}

The Casimir effect \cite{Casimir} is the dominant interaction between neutral 
plates separated by distances in the micron or submicron range.
The better and better control of this Casimir effect achieved over the last 
10 years \cite{Lamoreaux99,LambrechtPoincare,Milton04,ChenPRA04,DeccaAP05,OnofrioNJP06} hence 
opens new roads for the design of nanoelectromechanical systems (NEMS) \cite{capasso}. 

Casimir initially studied the simplest geometric configuration with two parallel 
plane plates large enough so that the theoretical analysis is simplified thanks
to the lateral translation symmetry \cite{LambrechtNJP06}. 
Except for a few cases \cite{OnofrioNJP06}, experiments are performed between a 
plane and a sphere \cite{ChenPRA04,DeccaAP05}, a geometry more easily mastered 
at distances in the micron or submicron range.
Force evaluations in this geometry are commonly obtained by using the so-called 
proximity-force approximation (PFA) \cite{Deriagin}, with the energy simply obtained 
by averaging the plane-plane expression over the distribution of local separation 
distances met in the plane-sphere geometry. 
It is commonly agreed that this approximation is valid when the radius $R$ of the 
sphere is much larger than the closest separation distance $L$ \cite{Jaffe04}. 
But a quantitative determination of its accuracy in plane-sphere experiments 
is still lacking, at least for the problem of experimental interest where  
electromagnetic fields are reflected on metals, the optical properties of
which must be accounted for at the distances met in the experiments. 
In contrast, results valid beyond the PFA have been reported 
for theoretical models involving scalar fields reflected on 
perfect boundary conditions \cite{Bulgac2006,Gies2006,Bordag2006}.   

At this point, we may stress that the situations which may be treated within 
the PFA correspond to a trivial interplay between geometry and Casimir effect
since the geometry is described by an averaging over the distribution of 
local distances. 
In contrast, the general case opens a far richer physics with a variety of 
stimulating theoretical predictions \cite{Balian77,Balian03,Balian04}, so that the exploration
of situations beyond reach of the PFA raises great expectations. 
The idea can already be tackled for the description of the effect of roughness
on the Casimir force. This description is commonly given within the PFA
\cite{Klimchitskaya99} valid only when the wavelengths associated with the plate 
deformation are large enough \cite{EPL2003,EPL2005,PRA2005}.
As the effect of roughness is only a small correction of the Casimir force,
and the characterization of the roughness state of the plates is not very
accurate, one can hardly expect quantitative theory-experiment
comparisons in this case.

Fortunately, there exists a geometry better suited to the aim
of an accurate theory-experiment comparison, namely that with parallel and
periodic corrugations imprinted on the metallic surfaces.
The Casimir force contains a lateral component besides the usual normal one, 
since lateral translation symmetry is broken here \cite{Golestanian}. 
The lateral Casimir force is smaller than the normal one, but it has
nevertheless already been measured in experiments \cite{Chen02}. 
It is easily computed within the PFA and has also been calculated beyond 
the PFA by using more elaborate theoretical methods. 
The lateral force has first been evaluated for perfect mirrors 
using a path-integral formulation in a perturbative \cite{Emig2003} 
or non perturbative approach \cite{Emig2005}. 
As expected, the PFA is found to be valid only in the limiting case 
where the corrugated surfaces are nearly plane for the vacuum fields
involved in the calculation of the Casimir energy.
When introducing the corrugation wavelength $\lambda_\C$ and the mean 
separation distance $L$, ones characterizes this limit as $\lambda_\C\gg L$. 

Now, the experiments have been performed with distances $L$ around 200nm 
at which it is essential to take the optical properties of the metals 
into account \cite{LambrechtEPJ00,GenetPRA00}.
A simple description of these optical properties is given by the plasma model 
introducing a further length scale, the plasma wavelength $\lambda_\P$,
with a typical value of 137nm for Gold plates.
A novel theoretical method has recently been presented \cite{PRL2006}
which allows one to calculate the lateral Casimir force for arbitrary 
relative values of the three length scales $L$, $\lambda_\C$ and $\lambda_\P$,
provided that the corrugation amplitudes $a_1$ and $a_2$ are smaller. 
The approach relies on scattering theory \cite{LambrechtNJP06} used in 
a perturbative expansion \cite{PRA2005} with respect to the corrugations.

In the present paper, we will first present the complete derivation 
of the results presented in \cite{PRL2006}.
We will write the lateral Casimir force at the order $\propto a_1 a_2$ 
for arbitrary relative values of $L$, $\lambda_\C$ and $\lambda_\P$.
Limiting cases will be obtained when these values obey some specific orderings. 
Some of them are already known, in particular the PFA limit and the perfect
reflection limit, and the known results are recovered as expected
in our calculations. 

We will conclude the paper by discussing the consequence of these results 
for comparison with existing measurements \cite{Chen02}.
These measurements were found to agree with the PFA computations, 
within the margins of experimental uncertainty \cite{Chen02}.
As this conclusion differs from that drawn from our results \cite{PRL2006},
there may be a potential concern for theory-experiment comparison, 
and the question will be discussed in detail at the end of this paper.
Let us recall that the corrugation amplitudes used in 
the experiments \cite{Chen02} were smaller, but not much smaller,
than the other length scales, so that the theoretical predictions drawn from
our perturbative expansion cannot be compared directly with the experimental 
results \cite{CommentPRL2007,ReplyPRL2007}. 
However, it seems unlikely 
 that the discrepancy demonstrated in the perturbative
regime will be exactly compensated by higher order terms in the perturbative expansion.
We will present a new result in the end of this paper which can be of relevance
for addressing the discrepancy.

\section{General outline and assumptions}

We first consider two parallel plane mirrors, M1 and M2, with corrugated surfaces
(the case of a plane and a sphere with corrugations will be studied later on).
The profiles of the mirrors M1 and M2 are defined by two functions $h_1(x,y)$ 
and $h_2(x,y)$ describing the local height with respect to mean planes 
$z_1=0$ and $z_2=L$. The mean planes are defined so that $h_1$ and $h_2$ have
null spatial averages, $L$ representing the mean distance between 
the two surfaces; $h_1$ and $h_2$ are both counted as positive when they correspond 
to separation decreases.

\begin{figure}[h]
\centering
\includegraphics[width=7cm]{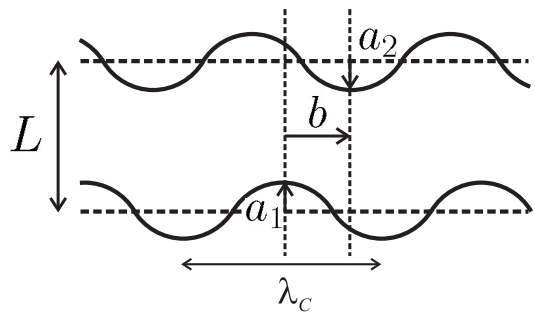}
\caption{Parallel corrugated surfaces, with $L$ representing the mean separation distance,
$a_1$ and $a_2$ the corrugation amplitudes and $b$ the lateral mismatch between
the crests. The corrugation are the smallest length
scales in the perturbative expansion used in the paper, they have been exaggerated for the sake
of a better visualization. }
\label{corrugation}
\end{figure}

We assume that uniaxial sinusoidal corrugations are imprinted on both plates with 
the same period $\lambda_\C$ and along the same direction, but with a spatial 
mismatch $b$ between the corrugations crests (see Fig.~\ref{corrugation})
\begin{eqnarray}
\label{h1h2}
&&h_1=a_1\,\cos(\kC x) \quad,\quad h_2=a_2\,\cos\left(\kC (x-b)\right) \nonumber\\
&&\kC=\frac{2\pi}{\lambda_\C}
\end{eqnarray}
$\lambda_\C$ is the corrugation wavelength and $\kC$ the corresponding wave vector. 
When a specific model of the optical response of the metallic mirrors will be needed, 
we will take the plasma model with the dielectric function 
\begin{equation}
\label{plasmamodel}
\epsilon(\omega) = 1-\frac{\omega _\P^2}{\omega^2} \quad,\quad 
\kP=\frac{2\pi}{\lambda_\P}=\frac{\omega_\P}{c}
\end{equation}
$\omega$ is the field frequency, $\omega_\P$ the plasma frequency, 
$\lambda_\P$ the plasma wavelength and $\kP$ the plasma wavevector. 

In the following, we will suppose that the corrugation amplitudes are smaller 
than the other length scales
\begin{equation}
\label{perturbat}
a_1, a_2 \ll \lambda_\C, \lambda_\P, L
\end{equation}
Within the range of validity of the PFA \cite{EPL2003}, the Casimir energy
in the presence of corrugations is simply obtained by adding over the mirrors' 
surfaces the contributions calculated with the local distances ${\cal L}$ 
\begin{equation} 
\label{PFA1}
{\cal E}_\PFA = \int \dd^2\br \frac{E_\PP\left({\cal L} (\br)\right)}A
\quad,\quad {\cal L} \equiv L-h_1-h_2
\end{equation}
Here $E_\PP/A$ is the Casimir energy per unit area calculated for plane and 
parallel plates, at the local separation distance ${\cal L}(\br)$. 
Using the condition (\ref{perturbat}) and expanding (\ref{PFA1})
up to second order, we find the lowest-order correction to energy
due to the presence of corrugations
\begin{eqnarray} 
\label{PFA2}
\delta {\cal E}_\PFA &=&  \frac 12 \frac{\partial^2 E_\PP}{\partial L^2} \; 
\int \frac{\dd^2\br}A \left(h_1+h_2\right) ^2 \\
&=& \frac 12 \frac{\partial^2 E_\PP}{\partial L^2} \; 
\left( \frac{a_1^2+a_2^2}2+a_1a_2\cos(\kC b) \right) \nonumber
\end{eqnarray}
As the energy corrections proportional to $a_1^2$ and $a_2^2$ 
do not depend on the lateral mismatch $b$, they do not contribute 
to the lateral force 
\begin{equation}
\label{FlatPFA}
F_\PFA^\lat = -\frac{\partial \delta {\cal E}_\PFA}{\partial b} = 
\frac 12 \frac{\partial^2 E_\PP}{\partial L^2} \; \kC a_1a_2\sin(\kC b) 
\end{equation}

In the following, the expression of the force will be extended
beyond the regime of validity of the PFA by using the general scattering 
approach presented in \cite{LambrechtNJP06}. 
Performing a perturbative expansion up to the second order in the corrugation 
amplitudes, we will be able to evaluate the lowest-order energy correction 
due to the corrugations, proportional as (\ref{PFA2}) to $a_1 a_2\cos(\kC b)$,
but with a different coefficient of proportionality in front of this quantity.
We will then find a lateral force 
\begin{equation}
\label{Flat}
F^\lat = \Gamma_\PP \,a_1 a_2\,\sin(\kC b)
\end{equation}
with $\Gamma_\PP $ a function of the three length scales 
$L$, $\lambda_\C$ and $\lambda_\P$.
The main aim of this paper is to obtain the explicit expression
of $\Gamma_\PP $ and to discuss various limiting cases.

First, we will discuss in some detail the PFA limit, that is the case
where (\ref{Flat}) can be reduced to (\ref{FlatPFA}).
This approximation was used for comparison with experiments in 
\cite{Chen02} and it has the advantage of being easily 
extended to higher orders in the corrugation amplitudes. 
However, as emphasized in \cite{EPL2003,EPL2005,PRA2005} for the
case of roughness and then in \cite{PRL2006} for the case of the lateral force, 
it can be valid only when the corrugation wavelength is larger than 
the other length scales ($\lambda_\C \gg L , \lambda_\P$). 
A second interesting limit corresponds to perfect reflection
of the mirrors. Then, the lateral force can be calculated 
in the path-integral theory developed in Ref.~\cite{Emig2003}, 
and higher order terms may also be calculated \cite{Emig2005}.
We will prove below that the path-integral theory provides us
with an independent test of our formalism since we recover 
its results in the limit $\lambda_\P \ll L, \lambda_\C$.

A third limiting case studied in the paper corresponds to the case
of rugged corrugations $\lambda_\C \ll L, \lambda_\P$.
This case corresponds to evaluations far beyond the PFA regime 
and is particularly interesting as it constitutes 
a non trivial interplay between
geometry and the Casimir effect \cite{Balian03,Balian04}.
It is also of great interest for applications to the 
configurations with nanometric corrugations. 
We will derive analytical expressions in this limit
and discuss the large deviation from PFA thus obtained.

In the present paper, we will restrict our attention on the specific
geometry where the corrugations of the two plates are aligned.
The scattering approach also allows one to consider the case where the 
corrugations are not aligned so that the Casimir energy depends on the angle 
between the two directions, which results in a torque \cite{EPL2006}.
The measurement of the Casimir torque with torsion balance techniques 
could be an alternative manner of testing the nontrivial geometry 
dependence of the Casimir effect. 
Other surface profiles than sinusoidal corrugations can also be 
considered, as long as the amplitudes remain smaller than the other 
length scales.

\section{The lateral Casimir force between corrugated plates}

In this section, we derive a general expression for the Casimir energy 
between corrugated plates up to second order in the corrugation amplitudes. 
We start from a general expression for the Casimir energy valid in the case of 
arbitrary nonspecular scattering \cite{LambrechtNJP06}. 
This expression contains second-order correction terms proportional to $a_1^2$
and $a_2^2$ which were studied in detail in \cite{PRA2005}, but do not
contribute to the lateral force.
Here we focus on the second-order correction terms proportional to $a_1a_2$,
which are responsible for the lateral force. 

The electromagnetic fields are developed over Fourier components labeled 
by the two-dimensional wave vector $\bk$ parallel to the $\br\equiv(x,y)$ 
plane and the polarization $p$ 
(transverse electric, TE, or transverse magnetic, TM). 
Their scattering upon the non plane mirrors is then described in terms of 
non specular reflection operators coupling different wave vectors and 
polarizations (more details in \cite{PRA2005}).
As a consequence of stationarity, scattering preserves the frequency $\omega$. 
There exist two relevant non specular reflection operators, 
${\cal R}_1$ and ${\cal R}_2$ which describe respectively the intracavity 
fields reflected by the mirrors M1 and M2 as functions of
the intracavity fields impinging these mirrors.

The Casimir energy between the two non plane mirrors is then written as an 
integral over imaginary frequencies $\xi=-i\omega$ \cite{LambrechtNJP06}
\begin{equation} 
\label{start}
{\cal E} = \hbar \int_0^{\infty} \frac{\dd \xi}{2\pi} \,{\rm Tr}
 \ln\left(1-{\cal R}_1(i\xi) e^{-{\cal K} L} {\cal R}_2(i\xi) e^{-{\cal K} L}\right)
\end{equation}
${\cal K}$ is a diagonal operator in the basis of plane waves $(\bk,p)$
with the diagonal elements given by $\kappa=\sqrt{\bk^2 + \xi^2}$. 
The trace in Eq.~(\ref{start}) is a sum over the plane waves defined as 
$\int \frac{\dd^2 \bk}{(2\pi)^2} \sum_{p} $.
In the simplest configuration with two plane parallel plates,
the reflection operator becomes diagonal with the diagonal elements 
given by the specular reflection coefficients $r_{j;p}(\bk,i\xi)$.
We thus recover from (\ref{start}) the known expression of the Casimir 
energy between plane parallel plates described by 
reflection amplitudes \cite{Jaekel91}
\begin{eqnarray}
\label{PP}
&&E_\PP = \hbar A \int_0^{\infty} \frac{\dd \xi}{2\pi} 
\int \frac{\dd^2 \bk}{(2\pi)^2} \sum_{p} 
\ln(d_p(\bk)) \nonumber\\
&&d_p(\bk) \equiv 1-r_{1;p}(\bk) r_{2;p}(\bk)\,e^{-2\kappa L}
\end{eqnarray}
The area $A$ of plates has been introduced as a substitute 
for the quantity $(2\pi)^2\,\delta^{(2)}({\bf 0})$.

The case of interest in this paper corresponds to corrugated plates. 
The reflection operators ${\cal R}_j$ thus contain zeroth-order 
contributions, denoted ${\cal R}_j^{(0)}$ and corresponding to plane 
surfaces, and non specular contributions induced by the corrugations
\begin{equation}
\label{refl}
{\cal R}_j = {\cal R}_j^{(0)} + \delta{\cal R}_j \quad,\quad
\delta{\cal R}_j = \delta{\cal R}_j^{(1)} + \delta{\cal R}_j^{(2)} + ...
\end{equation}  
Non specular operators $\delta{\cal R}_j$ have been expanded in powers 
of the Fourier transforms $H_j(\bk)$ of the surface profiles $h_j(\br)$. 
It is now straightforward to write a perturbative expansion of 
the Casimir energy (\ref{start}) in powers of the corrugation amplitudes.
The modification of the Casimir energy due to corrugations is read as
\begin{widetext}
\begin{equation}
\label{log}
\delta {\cal E}_\PP = \hbar \int_0^{\infty} \frac{\dd\xi}{2\pi}\,
 {\rm Tr} \ln \left[1-{\cal D}^{-1}\left(\delta{\cal R}_1 {\cal R}_2^{(0)} e^{-2{\cal K}L}+
{\cal R}_1^{(0)} e^{-{\cal K}L} \delta{\cal R}_2  e^{-{\cal K}L} 
+\delta{\cal R}_1 e^{-{\cal K}L} \delta{\cal R}_2 e^{-{\cal K}L}\right)
\right]
\end{equation}
%\end{widetext}
${\cal D}$ is a diagonal matrix with elements $d_p(\bk)$. 
We then collect the different orders in $H_j$.
The first-order terms, which represent the change of Casimir energy due to the mean
displacement of the mirrors, vanish thanks to the assumption $\langle h_j \rangle =0$.
The lowest-order corrections appearing at second order may be gathered 
in two categories, with square terms proportional to $H_1^2$ or $H_2^2$ 
(first line below) and cross terms proportional to $H_1H_2$ (second line),
%\begin{widetext}
\begin{eqnarray}
\delta {\cal E}_\PP &=& -\hbar \int_0^{\infty} \frac{\dd\xi}{2\pi}\,
 {\rm Tr}  \left[{\cal D}^{-1}\sum_{j=1}^2
\delta{\cal R}^{(2)}_j {\cal R}_{[j+1]}^{(0)} e^{-2{\cal K}L}+
\frac{1}{2}
\sum_{j=1}^2
\left( {\cal D}^{-1}
\delta{\cal R}^{(1)}_j
{\cal R}_{[j+1]}^{(0)} e^{-2{\cal K}L}\right)^2
\right]  \nonumber\\
&&- \frac{\hbar}{2} \int_0^{\infty} \frac{\dd\xi}{2\pi}\,
 {\rm Tr}
\sum_{j=1}^2
  \left({\cal D}^{-1}
\delta{\cal R}^{(1)}_j
{\cal D}^{-1}
 e^{-{\cal K}L}
 \delta{\cal R}^{(1)}_{[j+1]}
  e^{-{\cal K}L}
\right)
\label{Ec}
\end{eqnarray}
\end{widetext}
$[j+1]$ denotes a sum modulo 2, and thus substitutes each mirror by the other.
The second line has been simplified by noting that
$d_p^{-1}r_{1;p}r_{2;p}\,e^{-2\kappa L} = d_p^{-1} - 1$,
and then using the invariance of the trace under cyclic permutations.

The square terms (first line in Eq.\ref{Ec}) reproduce the roughness correction 
to the normal Casimir force, in full agreement with results reported in 
\cite{EPL2005,PRA2005}. As they do not contribute to the lateral force, 
we disregard these terms in the sequel of the paper.
In contrast, we focus our attention on the cross terms (second line in 
Eq.\ref{Ec}) which generate the lateral force evaluated up to the 
second order in the corrugation amplitudes.
We may incidentally note that the evaluation of the cross term is somewhat 
simpler than that of square terms, since the former is completely
determined by the first-order non specular reflection operators 
$\delta{\cal R}^{(1)}_j$.
The matrix elements of these operators are simply proportional 
to the Fourier components of the surface profiles, with the form
$R^{(1)}_{j;pp'}(\bk,\bk') H_j(\bk-\bk')$.
The coefficients $R^{(1)}_{j;pp'}$ depend on the optical properties 
of the mirror $j$ and  will be calculated below for metallic mirrors 
described by the plasma model.

Assuming that the two mirrors are made of the same medium, 
we write the cross contribution in (\ref{Ec}) as
\begin{eqnarray}
\label{G}
&&\delta {\cal E}_\PP^{\rm cross} = \int
\frac{\dd^2 \bk}{(2\pi)^2} \, {\cal G}(\bk) H_1(\bk) H_2(-\bk)
\nonumber \\
&&{\cal G}(\bk)=-\hbar \int_{0}^{\infty }\frac{\dd\xi }{2\pi }\int 
\frac{\dd^{2}\bk'}{(2\pi )^{2}}b_{\bk',\bk'-\bk}(\xi )   \\
&&
b_{\bk',\bk}=\sum_{p',p} \frac{e^{-(\kappa'+\kappa)L}
R^{(1)}_{p'p}(\bk',\bk) R^{(1)}_{pp'}(\bk,\bk') }
{d^{p}(\bk) d^{p'}(\bk')} \nonumber
\end{eqnarray}
For uni-axial sinusoidal corrugations (\ref{h1h2}),
this second-order cross contribution to energy is read as
\begin{equation}
\delta {\cal E}^{\rm cross}_{\rm PP} = \frac{A}{2}\,{\cal G}(\kC) \,a_1 a_2\, \cos(\kC b)
\end{equation}
Since ${\cal G}(k)$ is negative in (\ref{G}), the energy is minimized 
when the crests are facing each other ($b=0$ or a multiple of $\lambda_\C$).
Differentiating with respect to the lateral mismatch $b$, 
we obtain the expected expression (\ref{Flat}) of the lateral force 
with the function $\Gamma_\PP $ defined by
\begin{equation}
\label{GammaPP}
\Gamma_\PP \equiv \frac{A}{2}\,{\cal G}(\kC) \kC
\end{equation}
In the next two sections, we check out that the known cases of PFA and perfect reflection 
are recovered as appropriate limits of our more general scattering expression.
We then present more explicit results for the specific case of metallic mirrors.

\section{The Proximity Force Approximation}

The Proximity Force Approximation (PFA) has been written as Eq.~(\ref{PFA2}) above.
In the context of our calculations, we expect this expression to be recovered
in the limit of very smooth surfaces $\lambda_\C\rightarrow\infty$, that is 
precisely in the limit $\kC L, \kC\lambda_\P \ll 1$.
The question is thus whether or not the response
function ${\cal G}$ satisfies the condition
\begin{equation}
\label{PFTheorem}
A \lim_{k\rightarrow 0}{\cal G}(k) =
\frac{\partial^2 E_\PP}{\partial L^2} 
\end{equation}
The fact that this question has a positive answer can be proven 
in the context of our scattering formalism \cite{PRA2005}.

To this purpose, we have to study the 
specular limit of the non-specular scattering formalism. 
In this specular limit, the generalized reflection coefficients
show the following behavior
\begin{equation} 
\label{specular}
\lim_{\bk'\rightarrow\bk} R_{pp'}^{(1)}( \bk,\bk' ) = 2\kappa \,r_p\, \delta_{pp'}
\end{equation} 
This can be checked out on the explicit expressions given below for the plasma model
and is also true regardless of the model considered for the mirrors. 
As a matter of fact, $R_{j;pp'}(\bk,\bk)$ would give the correction of the Casimir
energy for a mean displacement of the mirrors $\langle h_j \rangle$
(should the latter not be supposed to vanish).
Hence it can be deduced from a global translation of the surface by a quantity 
$\langle h_j \rangle.$ For real frequencies, this amounts to an additional 
round-trip phase factor which is finally read as expression (\ref{specular}) 
when going to imaginary frequencies.   

{}From Eq.~(\ref{G}), we now deduce the specular limit of the response function ${\cal G}$
\begin{equation}
\lim_{k\rightarrow 0}{\cal G}(k) = - 4 \hbar  \int_{0}^{\infty }\frac{\dd\xi }{2\pi }\int 
\frac{\dd^{2}\bk}{(2\pi )^{2}} \sum_p \frac{e^{-2\kappa L} \kappa^2 r_p^2  }
{d_p^2}.
\end{equation}
This expression turns out to fit condition (\ref{PFTheorem}) when using the 
expression (\ref{PP}) of the Casimir energy $E_{\rm PP}$ evaluated between plane plates. 
The property (\ref{PFTheorem}) can be properly named as the ``Proximity Force Theorem'' with
a simplified expression indeed obtained in the secular limit of the more general 
non specular scattering theory. It holds in situations where specular 
scattering is sufficient to calculate the Casimir energy, that is also when the surfaces 
are approximately flat over distances of the order of $L$, since the main 
contributions to the Casimir effect come from wavelengths of the order of $L$.  

At this point, we want to emphasize that this discussion does by no means imply that
the limit (\ref{PFTheorem}) can be used as an approximation for arbitrary values of $\kC$.
In particular, the lateral force computed within the PFA grows linearly with the
corrugation wavelength $\kC$ whereas the scattering theory leads to a 
different behavior, eventually decreasing  
exponentially for large values of $\kC$.
In order to discuss deviation from PFA in an as clear as possible manner,
we introduce the ratio between the force calculated in the scattering and 
PFA theories, in both cases at the lowest order $\propto a_1a_2$,
\begin{equation}
\label{defrho}
\rho(\kC) = \frac{\Gamma_\PP}{\Gamma_\PP^\PFA}
= \frac{{\cal G}(\kC)}{{\cal G}(0)}
\end{equation} 
As a consequence of the discussions of the present section,
this ratio goes to unity in the PFA limit 
\begin{equation}
\label{PFArho}
\lim_{\kC\rightarrow 0} \rho(\kC) = 1
\end{equation}
The variation of this ratio $\rho$ for $\kC\neq0$, analyzed in the foregoing sections,
measures the inaccuracy of the PFA.

\section{The limit of perfect reflection
and the plasma model}

{}From now on, we take the plasma model (\ref{plasmamodel}) to describe metallic mirrors.
The ideal case of perfect reflection is expected to be reproduced at the limit 
of very small plasma wavelength $\lambda_\P \ll L, \lambda_\C$.
We first recall the results already known for ideal perfect
reflectors and which only depend on the parameters $L$ and $\lambda_\C$.
We then apply the general formalism derived above to the plasma model
and prove that the ideal case of perfect reflection is indeed reproduced
when $\lambda_\P$ is small enough.

The effect of geometry on the Casimir effect between perfectly reflecting mirrors
has been studied by various theoretical approaches 
(see \cite{Balian03,Balian04} for a review and \cite{Emig2006,Dalvit2006} for recent results).
The case of parallel plates with uni-axial corrugations has been 
studied in the perturbative second-order approximation \cite{Emig2003}
as well as in the nonperturbative case \cite{Emig2005}.
We restrict our attention here to the perturbative approximation,
the results of which are reproduced exactly by our scattering formalism
when the non specular reflection amplitudes known for perfect reflectors 
(see the Appendix C of Ref.~\cite{PRA2005}) are plugged into the expression
(\ref{G}) of the response function ${\cal G}(\kC)$. 

According to the general method discussed in the preceding section, 
we present the results in terms of the ratio (\ref{defrho}).
Changing the integration variables to $\gamma = \kappa L$ and $\gamma'=\kappa' L$,
we obtain the following expression 
\begin{eqnarray} 
\label{Gpr}
\rho = \frac{30}{\pi^4\kC L}&&\int_{0}^{\infty }\dd\gamma
\int_{\left| \gamma -\kC L\right| }^{\gamma +\kC L}\dd\gamma^\prime 
\,e^{-\gamma -\gamma ^{\prime }} \\
&&\times\frac{
 \frac{1}{4}\,[\gamma ^{2}+\gamma ^{\prime
2}-(\kC L)^{2}]^{2}+\gamma^{ 2}\,\gamma^{\prime 2}
}{\left( 1-e^{-2\gamma} \right) \left(
1-e^{-2\gamma ^{\prime }} \right) } \nonumber
\end{eqnarray}
The functional dependence of $\rho$ is very simple since it depends only
on the dimensionless variable $\kC L=2\pi L/\lambda_C$ which quantifies 
the smoothness of the surfaces on the scale determined by $L$.

\begin{figure}[h]
\centering
\includegraphics[width=7cm]{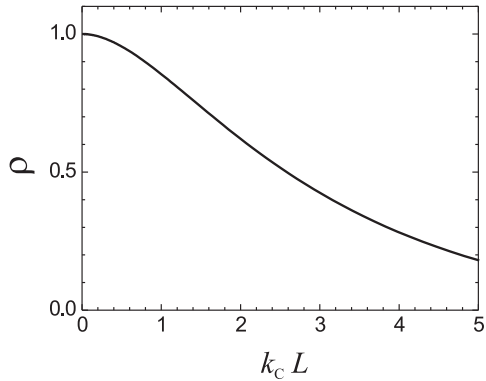}
\caption{Variation of $\rho$ versus the dimensionless variable $\kC L$ 
 for the ideal case of perfect reflection. }
\label{rho0}
\end{figure}

The PFA result $\rho\simeq 1$ is recovered at the limit of small
wavevectors $\kC\rightarrow 0$ (as predicted more generally by Eq.~\ref{PFArho}).
In the opposite limit of large values for $\kC L$, a completely
different behavior is obtained with an exponential decrease 
\begin{eqnarray}
\label{GprHk}
\rho =  \frac{30}{\pi^4} \left( \frac{(\kC L)^4 }{15}+(\kC L)^2
+3\kC L+ 3 \right) e^{-\kC L}
\end{eqnarray}
This expression is in full agreement with the results obtained in \cite{Emig2003}
and this is also the case for the numerical evaluation of (\ref{Gpr}) 
represented on Fig.~\ref{rho0}.
We note that the PFA result is reproduced with some accuracy only for very 
small values of the dimensionless variable $\kC L$.

We now turn to the study of metallic mirrors described by the plasma model 
where there is a third length scale, the plasma wavelength $\lambda_\P$. 
We recall the expressions of the specular Fresnel reflection amplitudes
\begin{eqnarray}
\label{Fresnel}
&&r_\TE(\bk,\xi) = \frac{\kappa-\kappa_t}{\kappa+\kappa_t}  
\quad,\quad
r_\TM(\bk,\xi)=\frac{\epsilon \kappa-\kappa _{t}}{%
\epsilon \kappa +\kappa _{t}} 
\\
&&\kappa _t\equiv\sqrt{\mathbf{k}^{2}+\epsilon \,\frac{{\xi }^{2}%
}{c^{2}}}=\sqrt{\kappa ^{2}+k_{\mathrm{P}}^{2}}
\quad,\quad
\epsilon \equiv 1+\frac{\omega _\P^2}{\xi ^2} \nonumber 
\end{eqnarray}
$\kappa_t$ denotes the imaginary part of the $z$ component of the wavevector
inside the metallic medium.
We also introduce the following shorthand notations 
\begin{eqnarray}
&&\beta = \frac{k}{\kappa} \quad,\quad 
\beta_t=\frac{k}{\kappa_t} \quad,\quad 
\mu_{\pm} = \frac{\kappa\pm \kappa_t}{1\pm \beta\beta_t}
\nonumber\\
&&h_p(\bk,\xi)=\frac{r_p(\bk,\xi)e^{-\kappa L}}{1-r_p(\bk,\xi)^2e^{-2\kappa L}}
\end{eqnarray}
We then use the nonspecular first-order reflection coefficients 
$R^{(1)}_{pp'}(\bk,\bk')$ computed in \cite{PRA2005}
with the help of the perturbation approach of \cite{Greffet}, 
based on the extinction theorem and the Rayleigh hypothesis \cite{Maradudin75,Agarwal77,Sanchez-Gil91}. 
When inserted into  Eq.~(\ref{G}), these expressions lead to 
\begin{eqnarray}
\label{bkkexp}
b_{\bk,\bk'} &=&  \sum_{\epsilon,\epsilon '=\,+,-}
\mu _{\epsilon } \mu_{\epsilon '}^\prime \\
&\times&\Bigl( h_\TE (\bk) h_\TE(\bk^\prime)   
C^{2}\left( 1+\epsilon \beta\beta _t\right) 
\left( 1+\epsilon ^\prime \beta^\prime\beta _t^\prime\right) 
\nonumber  \\
&& -\, h_{\mathrm{TE}}(\mathbf{k})h_{\mathrm{TM}}(\mathbf{k}%
^{\prime })  
S^{2}\left( 1+\epsilon \beta\beta _{t}\right)  \nonumber \\
&& - \, h_{\mathrm{TM}}(\mathbf{k})h_{\mathrm{TE}}(\mathbf{k}%
^{\prime })  S^{2}\left( 1+\epsilon ^{\prime }\beta
^{\prime }\beta _{t}^{\prime }\right)    \nonumber \\
&& +\, h_{\mathrm{TM}}(\mathbf{k})h_{\mathrm{TM}}(\mathbf{k}%
^{\prime })  \left( C+\epsilon \beta\beta
_{t}^{\prime }\right) \left( C+\epsilon ^{\prime }\beta ^{\prime
}\beta _{t}\right) \Bigr) \nonumber
\end{eqnarray}
The dependence on $\xi$ has been omitted, 
$C=\bk\cdot\bk'/(kk')$ and $S=\sqrt{1-C^2}$ represent the cosine and 
sine of the angle between the transverse vectors $\bk$ and $\bk'$.

\begin{figure}[h]
\centering
\includegraphics[width=7cm]{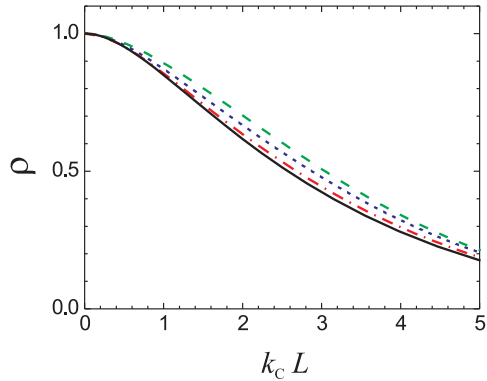}
\caption{Variation of $\rho$ versus the dimensionless variable $\kC L$ 
 for metallic mirrors described by the plasma model, 
for $\kP L=$1 (dashed line), 2.5 (dotted line), 5 (dashed-dotted line) 
and  10 (solid line) [colors online with respectively green, blue,
red and black lines]. }
\label{rho1}
\end{figure}

A number of interesting properties can be checked out analytically 
on the expression obtained by plugging (\ref{bkkexp}) into (\ref{G}). 
In particular, the PFA and perfectly-reflecting limits can again be 
recovered from this expression, by taking respectively 
$\kC\ll \kP, 1/L$ and $\kP\gg \kC, 1/L$. 
We chose here to illustrate the same properties by discussing
numerically integrated results. 
We plot $\rho$ as a function of $\kC L$, with fixed values of 
the second dimensionless quantity $\kP L$.
The values chosen for Fig.~\ref{rho1} ($\kP L=$ 1, 2.5, 5 and 10) 
correspond respectively to $L=$ 21.8, 54.5, 109, 218 nm 
when taking $\lambda_\P=137$ nm, which corresponds to gold-covered 
plates \cite{foot136}. 
In all cases, $\rho$ is smaller than unity and decreases when 
$\kC$ increases. As already seen for perfect mirrors, the accuracy
of the PFA is poorer and poorer for shorter corrugation wavelengths. 
At large values of $\kP L$, the curve obtained for the plasma model
tends towards the curve calculated for perfect mirrors, as expected
(compare the solid curve on Fig.~\ref{rho1} with that on Fig.~\ref{rho0}).
Otherwise, the result obtained for the plasma model differs from 
that for perfect mirrors. 

Let us give a few numbers here, with parameters 
$L=220\,$nm, $\lambda_\C=1.2\,\mu$m and $\lambda_\P=137\,$nm 
chosen to be close to the experimental figures of \cite{Chen02}. 
As stated in the preceding paragraph, this corresponds to a large value
$\kP L=2\pi L/\lambda_\P\simeq10$, so that the value of $\rho$ calculated 
for the plasma model approaches that obtained for perfect mirrors.
The precise values are $\rho = 0.814$ for the plasma model
and $\rho = 0.819$ for perfect mirrors.
The important point to be noticed here is that, contrarily to some
claims \cite{CommentPRL2007}, these values lie far
from the PFA expectation (which is simply $\rho = 1$). 
The discrepancy is larger in the experiment which employs a plane-sphere setup, 
as discussed in more detail below. 

To make this point completely clear, it is true that, for perfect mirrors, $\rho$ 
is determined exclusively by the quantity $\kC L$ (see discussions above). 
This is still approximately true for the plasma model as soon as the second
dimensional quantity $\kP L$ is large. 
Nevertheless, this property cannot be considered as universal since the
function $\Gamma_\PP$ is generally a function of the three variables $L$, $\kC$ and $\kP$,
and the dimensionless function $\rho$ a function of the two dimensionless 
variables $\kC L$ and $\kP L$.
To give an example, $L=55\,$nm and $\lambda_C=300\,$nm lead to the same value of $\kC L$
as in the preceding paragraph.
The values $\rho = 0.838$ found for the plasma model thus significantly differs from
$\rho = 0.819$ for perfect mirrors.
Even more striking illustrations of the dependence of $\rho$ on $\kP L$ will be given 
in the next section where we discuss the regime far beyond the PFA.

\section{Rugged corrugation limit}

We now consider the limit of short corrugation wavelengths 
$\lambda_\C \ll L, \lambda_\P$ which, as already noticed,
corresponds to a non trivial interplay between
geometry and the Casimir effect \cite{Balian03,Balian04} while being
of great interest for the problem of nanostructured surfaces. 

We have seen that $\rho$ decreases exponentially with $\kC L\gg 1,$ 
a property which can be given a simple explanation \cite{PRA2005,JPA2006}.
The cross contribution to corrugation energy (\ref{G}), evaluated at the lowest 
order, originates from intracavity propagation loops containing one non-specular 
reflection at each mirror. 
The first non-specular reflection changes the field momentum from $\bk'$ to $\bk,$ 
whereas the second changes the momentum back to its initial value
(see Eq.~\ref{bkkexp}). 
The exponential factors $\exp(-\kappa L)$ and $\exp(-\kappa' L)$ associated with
intracavity propagation (for imaginary frequencies) thus lead to the
following behavior in the rugged corrugation limit  
\begin{equation}
\label{rhobeta}
\rho \simeq \beta\,(\kC L)^{7/2}\, e^{-\kC L} \quad,\quad \lambda_\C \ll L, \lambda_\P
\end{equation}
$\beta$ is independent of $\kC L$ while remaining a function of $\kP L$.

Equation (\ref{rhobeta}) corresponds to a behavior very different from
that obtained for perfect mirrors (see Eq.~\ref{GprHk}).
This entails that $\rho$ takes very different values at the same $\kC L$ but
different $\kP L$, with
\begin{equation}
\frac{\rho(\kC L\gg 1,\kC L\gg \kP L)}{\rho(\kC L\gg 1,\kC L\ll \kP L)} 
\sim (\kC L)^{-1/2} \ll 1 
\end{equation}
In other words, the perfectly-reflecting model grossly overestimates the lateral 
Casimir force obtained in the limit of rugged corrugations.
Note that this occurs for a large separation distance $L$
though the latter condition is often carelessly associated with perfect reflection.
This is a serious limitation of the model of perfect reflection which, as already
emphasized, can only be trusted when $\lambda_\P$ is much smaller than
$\lambda_\C$ and $L$. 

\begin{figure}[h]
\centering
\includegraphics[width=7cm]{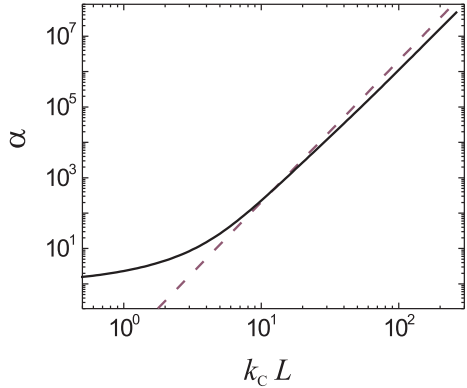}
\caption{Variation of $\alpha$ versus $\kC L$ (solid line); 
the high-$k$ limit for perfect reflectors is shown as the dashed line 
[colors online with respectively black and purple lines]. }
\label{alpha1}
\end{figure}

To propose a better visualization of this feature, we plot on Fig.~\ref{alpha1}
the quantity $\alpha\equiv\rho \exp(\kC L)$ for a large distance $L=1\,\mu$m
($\kP L=46$ with $\lambda_\P=$137nm), as a function of $\kC L$.
The result shown as the solid curve differs almost everywhere
from the result obtained for perfect mirrors 
$\alpha =  \frac{2}{\pi^4}\,(\kC L)^4$, shown as the dashed line. 
As $\kC L$ increases past $\kP L$, $\alpha$ stays below the result for perfect reflectors.

The exponential fall-off of $\rho$ at large $\kC L$ has to be contrasted 
with the linear increase obtained in previous papers 
for the similar function associated with roughness \cite{PRA2005}. 
Such a linear growth was resulting from the contribution of second-order 
non specular reflections which do not contribute to the lateral force. 
It follows that, whereas the PFA was underestimating
the roughness correction, it overestimates the lateral Casimir force.

\section{Comparison with experiments: the plane-sphere case}

We now consider the lateral Casimir force in the plane-sphere (PS) configuration, 
which corresponds to the experiments \cite{Chen02}.   
We derive the PS result from the plane-plane (PP) one by using the PFA
for treating the sphere curvature effect. The validity conditions required here 
are much more easily met than those needed for the corrugation effect. 
Indeed, experiments use spheres with radius $R$ of hundreds of microns, 
of the order of thousand times larger than the distance $L$. 
The condition $RL\gg\lambda_\C^2$ is also needed in order to treat
curvature and corrugation effects without taking any complicated
interdependence into account, and it is met  
with the experimental figures \cite{Chen02}.

We then use the PFA to deal with the sphere curvature effect, 
while accounting for deviations from the PFA for the corrugation effect. 
We obtain the cross energy correction $\delta {\cal E}_\PS^{\rm cross}$ in the PS case
from the one $\delta {\cal E}_\PP^{\rm cross}$ calculated for the PP case by integrating
over the distribution of distances generated by the sphere curvature
\begin{equation}
\delta {\cal E}_\PS^{\rm cross}(L,b) = \int_{L}^{\infty} 2\pi R \dd L'\,
\frac{\delta {\cal E}_\PP^{\rm cross}(L',b)}{A}
\end{equation} 
$b$ is the lateral mismatch of the parallel corrugations on the two plates
and $L$ the distance of closest approach. 
When differentiating with respect to $b$, one obtains the lateral Casimir force 
in the PS configuration as 
\begin{equation}
F_\PS^\lat = -\frac{\partial}{\partial b}\delta {\cal E}_\PS^\corr(L,b)= 
\int_{L}^{\infty} 2\pi R \dd L'\,\frac{ F_\PP^\lat(L',b)}{A}
\end{equation} 
This is then read exactly as expression (\ref{Flat}), with a function 
$\Gamma_\PS$ now written as (using Eq.\ref{GammaPP})
\begin{equation}
\label{GammaPS}
\Gamma_\PS = \int_{L}^{\infty} \pi \kC R \dd L'\, {\cal G}_{L^\prime}(\kC)
\end{equation}
Should the PFA be applied to the corrugation effect at all distances $L'>L$
involved in this integral (\ref{GammaPS}), ${\cal G}_{L^\prime}(\kC)$ 
would be replaced by ${\cal G}_{L^\prime}(0)$ (see Eq.~\ref{PFTheorem}),
leading to 
\begin{equation}\label{F0PFA}
\Gamma_\PS^\PFA =   \pi \kC R \frac{F_\PP}{A} 
\quad,\quad F_\PP\equiv -\frac{\dd E_\PP}{\dd L} 
\end{equation}
The accuracy of this approximation is well represented by the following ratio,
defined by analogy with (\ref{defrho}),
\begin{eqnarray}
\label{defrhoPS}
\rho_\PS(\kC) &=& \frac{\Gamma_\PS}{\Gamma_\PS^\PFA} \\
&=& \frac{1}{F_\PP(L)} \int_{L}^{\infty} \dd L' \, \frac{-\dd F_\PP(L')}
{\dd L'}\,\rho_{L^\prime}(\kC) \nonumber
\end{eqnarray} 
It is an immediate consequence of this expression
that the PFA accuracy will be worse in the PS configuration than in the PP one (for the
same $L$) since it is determined by the PP case with distances $L'$ larger than $L$.

\begin{figure}[h]
\centering
\includegraphics[width=7cm]{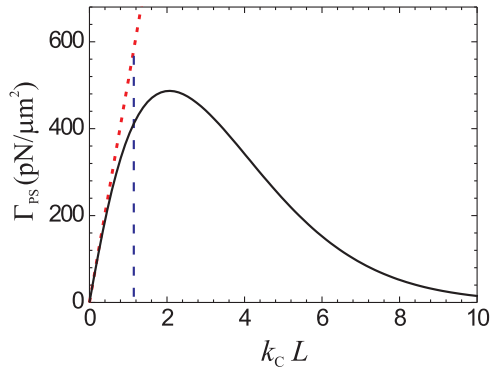}
\caption{Lateral force coefficient $\Gamma_\PS$ (force divided by $a_1a_2$) 
for the plane-sphere geometry, as a function of $\kC L$ 
($L$ and $\lambda_\P$ chosen to fit \cite{Chen02});
the  solid line is the result of scattering theory while 
the  dotted line corresponds to the PFA expression; 
the value $\kC L$ met in experiments is shown as the vertical 
dashed line [colors online with respectively black, red and blue lines].}
\label{FPSvsk}
\end{figure}

In order to stay closer to the experimental figures, we have chosen here
to illustrate these results by plotting on Fig.~\ref{FPSvsk} 
$\Gamma_\PS$ as a function of $\kC L$.
The parameters $L=220$nm and $\lambda_\P=$137nm are chosen to fit 
numbers of \cite{Chen02}.
The solid line representing the result of scattering theory has to be compared 
with the dotted line associated with the PFA.
The experimental corrugation wavelength $\kC\simeq5.2\,\mu$m$^{-1}$ 
is indicated by the vertical dashed line.
It is clear that the value thus attained by $\kC L=$1.14 is large enough 
to produce a significant inaccuracy of the PFA.
The precise numbers for this value are $\Gamma_\PS=421\,{\rm pN}/\mu{\rm m}^2$ and
$\Gamma_\PS^\PFA=585\,{\rm pN}/\mu{\rm m}^2$ respectively for scattering theory and PFA.
In other words, the scattering theory result is smaller by a factor $\rho_\PS\simeq 72\%$ 
than the PFA result with the same choice of parameters.
As expected, the inaccuracy of the PFA is significantly worse than it was 
in the PP case where $\rho=81\%$ was found (with the same choice of parameters).

Let us now discuss Fig.~\ref{FPSvsk} in a more general manner.
As previously, the PFA result (\ref{F0PFA}) grows linearly with $\kC$ and
this is also true for the scattering theory result at small $\kC L$.
When $\kC L$ increases, the scattering theory result begins to deviate from PFA
and eventually decays exponentially at large values of $\kC L$.
$\Gamma_\PS$ thus shows a peak value, found to lie at $\kC L=2.08$ 
(that is $\lambda_\C=665\,$nm when $L=220\,$nm).
For the PP setup, the peak position was found at $\kC L = 2.6$ and
the difference can again be explained from the fact that the PS force is 
an average of the PP result over $L'>L$.  

\begin{figure}[h]
\centering
\includegraphics[width=7cm]{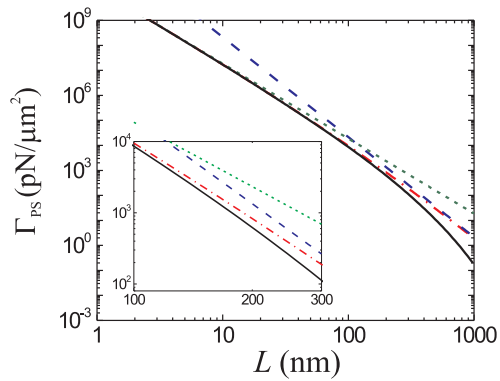}
\caption{Lateral force amplitude coefficient $\Gamma_\PS$ (force divided by $a_1a_2$) 
in the plane-sphere geometry, as a function of $L$, with  
$\lambda_\C$ and $\lambda_\P$ chosen to fit \cite{Chen02};
the curves correspond  to scattering theory (solid line), PFA (dotted-dashed line),
PFA combined with perfect reflection (dashed line) and plasmon (dotted line)
approximations [colors online with respectively black, red, blue and green
 lines]. }
\label{FPSvsL}
\end{figure}

We have also plotted on Fig.~\ref{FPSvsL} $\Gamma_\PS$ as a function of $L$ with
$\lambda_\C = 1.2\,\mu$m and $\lambda_\P = 137\,$nm fixed at their experimental values. 
The lateral Casimir force is found to decrease with the distance $L$.
The scattering theory result (solid line) corresponds to an exponential fall-off for $L > \lambda_\C$
whereas the PFA result (dot-dashed line) decays only as a power law, 
$L^{-3}$ in the plasmon regime ($L\ll \lambda_\P$, dotted line) 
and $L^{-4}$ in the perfectly-reflecting regime ($L\gg \lambda_\P$, dashed line). 
The scattering theory and PFA results agree at short separation distances $L\ll \lambda_\C$.
For very short distances $L\ll \lambda_\P<\lambda_\C$ furthermore, the force can be
deduced from the plasmon approximation \cite{GenetAFLB04,IntravaiaPRL}. 
The crossover between the $L^{-3}$ behavior in this range and the exponential decay 
at long distances occurs in the region around a few hundred nanometers, which is 
magnified in the inset of Fig.~\ref{FPSvsL}. 
Note that this is also the range tested experimentally, 
which we will discuss now.

We come now to a discussion of the number $\rho_\PS\simeq 72\%$, 
which measures the inaccuracy of PFA, and 
points at a potential concern for the theory-experiment comparison.
We recall that the experiments were performed with corrugation amplitudes 
smaller, but not much smaller than the other length scales 
(\cite{Chen02} report $a_1= 59\,$nm, $a_2= 8\,$nm,
to be compared to $\lambda_P=137\,$nm, $L=220\,$nm, $\lambda_C=1.2\,\mu$m).
As already discussed in \cite{PRL2006}, this point made a direct comparison 
between experiments and our perturbative scattering theory impossible, and
thus pushed us to try to discard the higher orders 
contribution to the PFA calculation of \cite{Chen02} 
in order to perform an indirect comparison. 
As higher order corrections have not yet been estimated within
scattering theory, we will not pursue this line of reasoning further.

Here we want to emphasize the comparison between results obtained within
PFA and beyond PFA for the figures of experimental interest.
When restricting the attention to calculations up to the second order,
this comparison is precisely characterized by the number $\rho_\PS\simeq\,72\%$ 
calculated for the parameters favored in \cite{CommentPRL2007}. 
In our opinion, this discrepancy ($\sim$ 28\%) does not lie so far from
the margins of experimental uncertainty (0.32$\pm $0.077pN in \cite{CommentPRL2007}), 
which correspond to a relative accuracy of $\pm 24\%$. 
In contrast to this opinion, 
the concern was made more acute by a Comment \cite{CommentPRL2007} 
which fabricated a larger discrepancy by comparing two numbers 
which are {\it not} to be compared (and which we did {\it not} compare),
namely the perturbative result beyond the PFA and the non perturbative 
result within the PFA.
We have already explained our point of view in a Reply \cite{ReplyPRL2007}, 
and now present new discussions of the issue
in the sequel of the section. 

An interesting way of addressing the issue is suggested by a close
scrutiny of Fig.~\ref{FPSvsL}. If we fit the result of scattering theory
to a power law within the interval of experimental interest,
we obtain a law $\propto L^{-4.1}$ in agreement with experiment \cite{Chen02}. 
This coincidence is also apparent in the fact that the scattering theory curve
is roughly parallel in this distance interval to the law obtained by combining PFA 
and the model of perfect reflection (see the inset of Fig.~\ref{FPSvsL}). 
This suggests that the difference between these two curves can be confused with
a poor determination of the absolute value of the distance $L$ between the mirrors.
As far as theory is concerned, $L$ is precisely defined as the 
{\it mean} separation distance between the corrugated surfaces, 
such that the corrugation profiles have zero mean values 
(see Fig.~\ref{corrugation}).
Precise measurements of {\it variations} of $L$ were reported in 
\cite{Chen02}, but the absolute determination of its value 
was by far more difficult, in a geometrical configuration where 
not so small corrugations are facing each other on plane and spherical surfaces.

\begin{figure}[h]
\centering
\includegraphics[width=7cm]{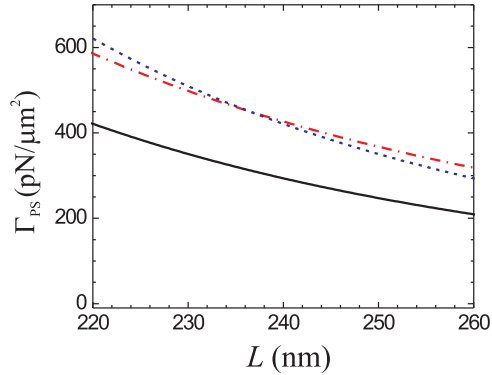}
\caption{Same conventions as in Fig.~\ref{FPSvsk}, with curves restricted to the range 
of distances from 220 to 260 nm; the  solid and  dashed-dotted lines represent 
the scattering theory and PFA results calculated for the value of $L$ indicated by the abscissa;
the  dotted line is the scattering result computed with an offset of 20 nm for $L$
[colors online with respectively black, red and blue lines].}
\label{offset}
\end{figure}

In order to test the idea that the difference between scattering theory and PFA 
could be confused with an offset in the determination of the absolute distance $L$,
we plot on Fig.~\ref{offset} three curves in the range from 220 to 260 nm.
The solid and dotted lines represent the scattering theory and PFA results calculated
with the value of $L$ supposed to be ideally determined; the third, dashed-dotted, line 
describes the scattering theory result with an offset of 20 nm for the distance $L$.
The important point is that the scattering theory with an offset can easily be confused
with the PFA without it. More precisely, the difference between the two curves is
certainly within the margins of experimental uncertainty, the magnitude of which is
of the order of the initial difference between the scattering theory and PFA results.

This means that the difference between scattering theory and PFA can have been 
confused with an offset of 20 nm in the distance measurement.
Note that this value is well below the larger corrugation amplitude (59 nm) in 
the experiment. It is therefore consistent with the already discussed 
experimental difficulty for localizing the position of the 
abstract reference planes at a scale smaller than the corrugation amplitudes. 
Note also that, if the absolute determination of the distance has been helped
by a global fit of the experimental results to the theory, the offset should
have been automatically produced by the fitting procedure.

\section{Conclusion} 

We have studied the lateral Casimir force arising between two corrugated metallic plates,
using scattering theory in a perturbative expansion valid when the corrugation amplitudes 
are smaller than the other length scales $L$, $\lambda_\C$ and $\lambda_\P$. 
We have shown that the Proximity Force Approximation (PFA) is recovered at the limit
of smooth plates $\lambda_\C\gg L,\lambda_\P$ and we have also obtained 
an expression for the lateral force in the opposite limit of
rugged corrugations $\lambda_\C\ll L,\lambda_\P$.
We have reproduced the results known for perfect mirrors
when $\lambda_\P\ll\lambda_\C,L$ and have also
given expressions valid when this is not the case. 

As the perturbation conditions $a_1, a_2 \ll L, \lambda_\C, \lambda_\P$ are not met
in the experiment \cite{Chen02}, the comparison of scattering theory with 
measurements of the lateral force will require more work.
Progress on this question could be achieved by calculating higher order 
corrections for metallic mirrors beyond the PFA.
These corrections would affect the numbers given in the present paper, but it
is unlikely that they could compensate exactly the deviation from PFA which has
been demonstrated in the perturbative theory. 

Progress could alternatively come from further experiments.
It would be very useful to have experiments with
small corrugation amplitudes
attaining the domain of validity of the perturbative theory.
Other relevant improvements would be to increase the experimental accuracy,
in order to be able to distinguish more easily between alternative
theories, and if possible to measure the absolute distance in a more 
reliable manner, in order to get rid of the offset confusion. 
Of course, this program raises serious experimental challenges, 
given the minuteness of the lateral force effect. 
But the reward would be remarkable with a potential experimental
access to a configuration where arises a nontrivial interplay 
between geometry and the Casimir effect, that is beyond the PFA.

\acknowledgments

R.B.R. and P.A.M.N. thank FAPERJ, CNPq
and Institutos do Mil\^enio de Informa\c c\~ao Qu\^antica e Nanoci\^encias
for financial support. A.L. acknowledges partial financial support 
by the European Contract STRP 12142 NANOCASE.

\end{document}